\documentclass[sigconf]{acmart}
\AtBeginDocument{%
  }

\setcopyright{acmlicensed}
\copyrightyear{2018}
\acmYear{2018}
\acmDOI{XXXXXXX.XXXXXXX}
\acmConference[Conference acronym 'XX]{Make sure to enter the correct
  conference title from your rights confirmation email}{June 03--05,
  2018}{Woodstock, NY}
\acmISBN{978-1-4503-XXXX-X/2018/06}

\begin{document}

\title{From Raw IDs to Semantic Planning: How Recommender Systems Utilize Information at Scale}

\author{Changhong Jin}
\email{changhong.jin@ucd.ie}
\affiliation{%
  \institution{University College Dublin}
  \city{Dublin}
  \country{Ireland}
}

\author{Shiqiu Yang}
\email{shiqiu.yang@ucdconnect.ie}
\affiliation{%
  \institution{University College Dublin}
  \city{Dublin}
  \country{Ireland}
}

\author{Roger Zhe Li}
\email{roger.zhe.li@huawei.com}
\affiliation{%
  \institution{Huawei Ireland Research Centre}
  \city{Dublin}
  \country{Ireland}
}

\author{Yingjie Niu}
\email{yingjie.niu@huawei.com}
\affiliation{%
  \institution{Huawei Ireland Research Centre}
  \city{Dublin}
  \country{Ireland}
}

\author{Aghiles Salah}
\email{aghiles.salah@h-partners.com}
\affiliation{%
  \institution{Huawei Ireland Research Centre}
  \city{Dublin}
  \country{Ireland}
}

\author{Mete Sertkan}
\email{mete.sertkan@h-partners.com}
\affiliation{%
  \institution{Huawei Ireland Research Centre}
  \city{Dublin}
  \country{Ireland}
}

\author{Zheng Ju}
\email{zheng.ju@ucd.ie}
\affiliation{%
  \institution{University College Dublin}
  \city{Dublin}
  \country{Ireland}
}

\author{Xingsheng Guo}
\email{xingsheng.guo1@huawei-partners.com}
\affiliation{%
  \institution{Huawei Ireland Research Centre}
  \city{Dublin}
  \country{Ireland}
}

\author{Huifeng Guo}
\email{huifeng.guo@huawei.com}
\affiliation{%
  \institution{Huawei Technologies}
  \city{Shenzhen}
  \country{China}
}

\author{Ruihai Dong}
\email{ruihai.dong@ucd.ie}
\affiliation{%
  \institution{University College Dublin}
  \city{Dublin}
  \country{Ireland}
}

\author{Barry Smyth}
\email{barry.smyth@ucd.ie}
\affiliation{%
  \institution{University College Dublin}
  \city{Dublin}
  \country{Ireland}
}

\renewcommand{\shortauthors}{Trovato et al.}

\begin{abstract}
The evolution of recommender systems can be explored by asking how they utilize information at scale. 
Throughout most of the historical period under consideration during the past two decades, industrial systems have relied on raw IDs, which are discrete, globally unique, and semantically opaque identifiers that enable exact lookup, logging, and item-specific memorization at scale. 
Over time, however, recommender systems have sought to utilize richer sources of information, including item content, context, multimodal signals, and cross-domain structure.
This development has led to a new stage in which part of such information is no longer used solely as auxiliary features around item identity, but is increasingly encapsulated in semantic IDs that provide a more structured, model-facing form of identity. 
We argue that this shift goes beyond the rise of generative recommendation over traditional methods. Indeed, it reflects a broader evolution in how recommender systems utilize information under industrial-scale constraints. 
This paper looks at the past, present, and future to examine three connected questions: why raw IDs dominated the early development of recommender systems, why semantic information is increasingly being encapsulated in IDs today, and what may come next once recommendations move beyond semantic retrieval. 
In particular, we introduce semantic planning as a possible future direction in which the system first predicts the semantic target of the next exposure, and only then instantiates that target as a specific item or generated creative. 
We further argue that such a shift may require changes not only in model design but also in evaluation and in the way recommender systems coordinate the objectives of users, platforms, and providers.
\end{abstract}




\maketitle

\section{Introduction}
Many real-world recommender systems, especially industrial ones, are multi-stakeholder systems~\cite{DBLP:journals/umuai/AbdollahpouriAB20}. 
Their primary role is to help users navigate large choice spaces and information overload~\cite{jannach2010recommender}, but in practice, they should also account for the commercial and strategic objectives of other stakeholders, including platform providers, online product providers, and advertisers~\cite{li2023metric, jannach2023survey}. 
In such settings, recommendation is not only a problem of predicting what a user may like; it is also about optimizing the decisions that need to be made by a business operating at scale.

From this perspective, a useful lens on the past two decades of recommender systems is \textbf{how they utilize information at scale}. 
In this context, the term \textit{information} extends well beyond historical interaction logs. 
The scope of this field encompasses item content~\cite{DBLP:reference/rsh/LopsGS11, DBLP:conf/kdd/WangB11, DBLP:conf/kdd/WangWY15, DBLP:conf/aaai/HeM16}, contextual signals~\cite{DBLP:journals/aim/AdomaviciusMRT11, DBLP:conf/ijcai/SalahL18}, multi-view behaviors~\cite{DBLP:conf/pakdd/KimKLJS25}, cross-domain structure~\cite{DBLP:journals/tois/ZangZLZY23, DBLP:conf/ijcai/YuanYB19, DBLP:conf/recsys/SalahTL21}, and increasingly heterogeneous forms of metadata and feedback~\cite{DBLP:journals/tkde/ShiLZSY17, DBLP:journals/aiopen/LiuSYLY22}. 
Many influential developments in recommender systems can be viewed as attempts to incorporate richer and richer signals. 
In particular, the field has repeatedly sought representations that improve generalization while remaining compatible with large vocabularies, growing embedding tables, and production-serving constraints~\cite{DBLP:conf/recsys/Cheng0HSCAACCIA16, DBLP:journals/corr/abs-1906-00091, DBLP:conf/recsys/ShtoffVHSRB24, DBLP:conf/recsys/FukumotoSSTSMX25}.

In this paper, we argue that one useful way to understand recent recommender-system evolution is as a flow from \emph{Raw IDs}, to richer \emph{semantic information} around IDs, and increasingly to \emph{semantic IDs}. 
Early industrial recommender systems therefore relied heavily on raw IDs as stable categorical inputs and operational keys for lookup, logging, ranking, and memorization at scale~\cite{DBLP:conf/www/HeLZNHC17, DBLP:conf/sigir/0001DWLZ020}. 
Later systems enriched these pipelines with statistical features~\cite{DBLP:conf/ijcai/GuoTYLH17, DBLP:conf/kdd/WangFFW17}, reviews~\cite{DBLP:conf/wsdm/ZhengNY17}, visual signals~\cite{DBLP:conf/aaai/HeM16}, contextual features~\cite{DBLP:conf/recsys/Cheng0HSCAACCIA16, DBLP:journals/tmis/UngerTL20}, and sequential behavior~\cite{DBLP:conf/kdd/ZhouZSFZMYJLG18}, while largely keeping raw IDs as the identity layer. More recently, semantic-ID and tokenization work has explored incorporating part of this richer information into the IDs themselves, in response to large, dynamic catalogs, embedding table growth, and the need for more unified model interfaces toward a "One-Model-Fits-All" effect~\cite{DBLP:conf/recsys/PenhaDNPTVLLHFB25, DBLP:conf/nips/RajputMSKVHHT0S23}. What is changing today is therefore not only that recommender systems use richer signals, but that part of their semantic structure is increasingly being moved into the identity layer itself.

This development trajectory can be understood as a sequence of progressively wider unifications. The first stage brought unification within recommendation: semantic IDs allow items from the same domain to share structural identity, enabling better generalization and cold-start handling. 
The second stage extended this to unification across domains and modalities, as semantic IDs provide a more common substrate for cross-domain transfer and multimodal alignment. 
The third, ongoing stage concerns the unification of tasks: semantic IDs are increasingly shared between search and recommendation. 
Yet all three stages remain primarily focused on better user-centered matching over existing items. The critical frontier that none of them fully addresses is explicit multi-stakeholder decision-making, i.e., how to reason over the potentially conflicting objectives of users, platforms, and providers within a single recommendation decision.
Under this circumstance, one possible direction is \emph{semantic planning}, which may emerge as the next stage. 
Once a recommender system can represent not only what an item is, but also what the next exposure is intended to accomplish, recommendation is no longer limited to selecting the nearest available item.
The system can instead first determine the semantic target of the next exposure --- a representation of what it should accomplish in terms of user need, platform objective, and provider content --- and only then instantiate that target as a specific item, or generated creative. Semantic planning is thus the process; the semantic target is its intermediate output. This paper argues that the shift from raw IDs to semantic IDs sets the stage for this transition, and that semantic planning represents its most consequential implication.



\section{Past: Raw IDs as Operational Anchors}
Raw IDs are discrete, globally unique, and semantically opaque identifiers used to index items in recommender systems. 
In industrial pipelines, such IDs usually function as stable database keys across retrieval, lookup, logging, storage, and serving~\cite{DBLP:journals/corr/abs-1906-00091}.
They allow a consistent track \& trace on the same item across heterogeneous system components, even when the models, features, and business rules around that item change.

Raw IDs became central because they resolve two practical problems. 
Firstly, they provide a stable method for connecting catalog entities to the numerous operations required by an industrial recommender system under high concurrency. 
Crucially, as highlighted in the seminal work on ML technical debt~\cite{DBLP:conf/nips/SculleyHGDPECYC15}, industrial systems often suffer from "implicit coupling", in which multiple infrastructure components rely on these stable anchors for cross-service synchronization. 
In this sense, raw IDs are not merely a modeling choice, but an architectural necessity to maintain stability in complex and high-concurrency environments.
Secondly, they provide models with a direct way to accumulate and aggregate item-specific behavioral evidence. This matters because semantically similar items can still develop different interaction histories and business value. 
While two items might share identical semantic features, their real-world trajectories, such as price, availability, popularity, or commercial value, may still diverge. 
Historically, raw IDs provided the most direct mechanism for memorizing such behavioral differences without requiring them to be explained through semantic features~\cite{DBLP:conf/recsys/Cheng0HSCAACCIA16}. 
However, this memorization comes with a structural limitation. 
By treating each item as an atomic, opaque unit, the system is forced to learn each item's nuances from scratch. 
While this approach provides high fidelity for head items, it also suffers from several limitations, including cold-start~\cite{DBLP:journals/corr/abs-2501-01945} and the growing memory and serving overhead caused by increasingly large embedding tables~\cite{DBLP:journals/corr/abs-1906-00091}.


As recommender systems evolved, richer semantic signals such as text and images were increasingly incorporated into models, but item identity itself remained largely Raw-ID-based. The issue, then, was not a lack of semantic information, but that it remained largely decoupled from the identity layer. This tension between representing items through semantic features and identifying them through arbitrary indices motivates the shift toward semantic IDs. The goal of this shift is to move from unconstrained memorization toward a structured identity that more directly bridges content and behavior~\cite{DBLP:conf/nips/RajputMSKVHHT0S23}.


\section{Present: From Semantic Information to Semantic IDs}

A recommender system can make extensive use of semantic information while keeping item identity organized in largely the same way.
Reviews, images, contextual features, and behavioral signals may all enrich the model, while raw IDs still remain the primary identifiers~\cite{DBLP:conf/recsys/Cheng0HSCAACCIA16, DBLP:conf/recsys/TruongSL21}. 
Semantic IDs mark a further step: \textbf{part of that information is encapsulated in the identifiers themselves, turning item identity into a more structured, model-facing object~\cite{DBLP:conf/nips/RajputMSKVHHT0S23}}. 
Their significance, therefore, lies not in the mere availability of semantic signals, but in reorganizing part of those signals into the ID itself.
The key advantage over embedding-based alignment is at the model interface level: semantic IDs allow us to decouple embedding alignment from task-specific models, enabling more flexible integration of heterogeneous item representations. As discrete token sequences, semantic IDs are self-contained and can be directly consumed by any model operating over a shared vocabulary—whether generative or retrieval-oriented—without additional cross-space alignment~\cite{DBLP:conf/nips/Tay00NBM000GSCM22, DBLP:conf/www/HouHMZ23}. This flexibility is particularly consequential in settings where useful structure spans multiple domains, modalities, or task boundaries, precisely the long-standing challenges that have motivated the present stage of recommender system development.

Traditionally, the model attempts to transfer knowledge by aligning or disentangling shared and domain-specific representations in continuous embedding spaces~\cite{DBLP:journals/tois/ZangZLZY23, DBLP:conf/nips/MaZ0Y019, DBLP:conf/ijcai/ManSJC17}. 
Recent work suggests that part of the cross-domain structure can be encapsulated in the IDs themselves. 
This allows items from different domains to be related through a more unified identity space~\cite{DBLP:conf/www/HouHMZ23}, where domain-agnostic kernels and domain-specific nuances are effectively unified~\cite{DBLP:journals/corr/abs-2507-12871} or factorized~\cite{DBLP:conf/aaai/HuLW26} within the ID structure itself. 

A similar shift is taking place in multimodal recommendation. Images, text, and other modalities have long been used as side information, especially in sparse and cold-start settings. 
At present, however, the goal is increasingly to align and encode part of that multimodal structure into semantic IDs. By doing so, multimodal signals are organized into the identifiers themselves, making them natively accessible to retrieval-, ranking-, and generation-oriented models~\cite{DBLP:journals/corr/abs-2503-23333, DBLP:conf/cikm/0006P0000LJ025, DBLP:conf/wsdm/XuZL0X0H0Z026, DBLP:journals/corr/abs-2602-10445}. 

Additionally, recent progress has further incorporated collaborative signals into semantic IDs, enabling them to carry information from both users and items~\cite{DBLP:conf/recsys/Bao0LZSFC25, DBLP:conf/www/HeHHKMMSTWWYBCC26}. 
This trend also facilitates the use of semantic IDs in the growing unification of search and recommendation. 
Recent progress in the community makes this connection more direct by enabling more shared item representations, generation mechanisms, and retrieval interfaces across the two tasks~\cite{DBLP:conf/recsys/PenhaDNPTVLLHFB25, DBLP:conf/recsys/ShiX0Z00Y25, DBLP:journals/corr/abs-2601-09496}. 
This does not mean that search and recommendation are already fully unified. 
It suggests that semantic IDs make it easier to share item representations and retrieval interfaces.

Taken together, these developments share a common character: they unify representation and retrieval around better matching of users to existing items. Each makes it easier for a system to find and present what a user is likely to want. Yet these benefits often come with added complexity in ID stability and lifecycle management and, more importantly, leave the multi-stakeholder structure of industrial recommendation largely untouched. Users, platforms, and providers each have distinct and often conflicting objectives, and none of the above unifications provides a principled mechanism for reasoning over them within a single recommendation decision. This is the boundary of the present stage. The future begins not when better semantic representations of items become available, but when the system must reason explicitly about \textbf{what the next exposure should accomplish under competing user, platform, and provider objectives}.

\section{Future: Semantic Planning and Implications}

The shift from semantic retrieval to semantic planning has implications that extend beyond model design and may open up new R\&D avenues. We examine three directions: the introduction of an intermediate planning layer before item selection, the consequent flexibility in how semantic targets are instantiated, and the changes that may be brought to evaluation and industrial deployment.

\subsection{From Semantic Retrieval to Semantic Planning}

Most current semantic recommender systems remain, at their core, retrieval systems. The generated object is usually an identifier that resolves to a specific item already in the catalog; "generation" is employed principally as a retrieval mechanism~\cite{DBLP:journals/corr/abs-2304-10149}. 
However different their architectures, they are all designed to answer the same question: \textbf{which existing item should be shown next?} This framing has a structural consequence: the system's only degree of freedom is the choice among available items, and its only explicit objective is user-item matching. The multi-stakeholder purpose of the next exposure—what it should accomplish for the user, the platform, and the provider—remains implicit, absorbed into item scores rather than reasoned about directly.


Consider a simple hotel recommendation scenario. 
A user is planning a weekend trip and has recently browsed several hotels. 
Their behavior may imply an ideal semantic target, such as accommodation in the city center that offers safety assurances, a good breakfast, a gym, and the option of free cancellation.
Suppose, however, that no existing hotel in the catalog satisfies all of these conditions. The closest available option may be a centrally located, well-reviewed hotel that includes breakfast but without a gym or free cancellation. A retrieval-oriented system may still recommend this hotel because it is the nearest available match. The user may click the item, read the reviews, and then leave the page after noticing that the cancellation policy is missing. This response is more than a preference among existing hotels. It reveals an existent semantic requirement that the current inventory has failed to satisfy~\cite{DBLP:conf/nips/LiuCBSZWW24}.

\begin{figure}[th]
    \centering
    \includegraphics[width=.92\linewidth]{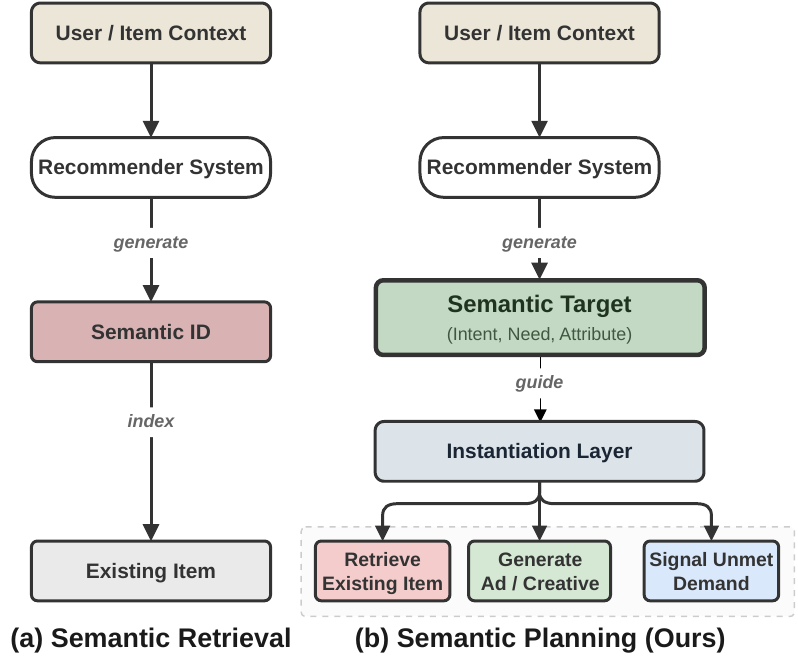}
    \caption{\small Paradigm shift from semantic retrieval to semantic planning. (a) Semantic Retrieval maps the context directly to a semantic ID to fetch an existing catalog item; (b) Semantic Planning separates the semantic target from the instantiation.}
    \Description{}
    \label{fig:ppf}
\end{figure}

A planning-oriented recommender would introduce an intermediate decision layer before item selection. Rather than directly asking which item to be shown, it would first ask: \textbf{what should the next exposure accomplish?} 
Semantic IDs may make this intermediate layer more tractable: raw IDs provide little semantic structure that can be transferred across items, while continuous embeddings are harder to use as an explicit and legible vocabulary for planning targets. By contrast, structured semantic IDs can provide a more grounded representation through which such targets can be expressed and through which failures of current inventory to satisfy them can be surfaced more clearly.
In the hotel example, the answer may be that the user no longer needs further destination discovery and instead needs reassurance during the booking process. 
The next exposure may therefore need to emphasize flexibility, reduce perceived risk, or help the user compare low-commitment options. This objective is grounded in the user’s current need, but it may also be shaped by platform goals and provider-side content. 
Semantic planning, therefore, makes the purpose of the next exposure explicit before deciding how that purpose should be instantiated. This distinction separates semantic planning from semantic retrieval as depicted in Figure~\ref{fig:ppf}.

Semantic retrieval maps the current user state to an identifier that resolves to a specific item, while semantic planning aligns the current context with a semantic target specifying the role, intent, or desired properties of the subsequent exposure.
The system will only instantiate the target as a hotel, an offer, a sponsored message, or another form of content once the target has been determined.
This also differs from conversational recommendation \cite{DBLP:journals/corr/abs-2308-10053}, which requires explicit dialogue to surface needs, and from intent modeling, which infers user goals but maps them directly to item scores. 
In semantic planning, the intermediate target is explicitly represented and optimized as a multi-stakeholder decision variable, not merely a latent state before item selection. 
Within this framework, the semantic structure learned from item IDs may serve as the basis for an intermediate control layer~\cite{DBLP:conf/naacl/WangJCYZCFLHY24}. Such a layer can represent targets such as flexible city-center accommodation, low-risk options for hesitant bookers, or family-friendly alternatives near public transport before mapping them to catalog items. It is important to note that separating the target from the item also enables the system to identify instances in which the current inventory cannot satisfy the inferred need~\cite{DBLP:journals/corr/abs-2304-03516}.

\subsection{From Semantic Targets to Instantiation}
Once recommender systems can represent what the next exposure should accomplish, recommendations are no longer limited to selecting the closest existing item. The system can instantiate the semantic target through items, messages, offers, or generated creatives, making recommendations a more flexible mechanism for coordinating users, providers, and platforms.

The most direct form of instantiation remains item selection. Once the system predicts a semantic target, it can retrieve the item or offer that best realizes that target. In the hotel example, a target such as low-risk city-center accommodation for hesitant bookers may be instantiated as a hotel with free cancellation, a flexible booking package, or a sponsored listing that clearly highlights refund protection. This differs from ordinary retrieval because the system no longer just does user-item matching. It first makes explicit what the next exposure is intended to achieve, then searches the item, offer, or message that best serves that purpose. Even when the final output is still an item, the intermediate target gives the system a clearer basis for comparing alternative instantiations.

A more flexible form of instantiation becomes possible when the system can generate or adapt the exposure's content. This is especially natural in advertising, where the effectiveness of a recommendation depends not only on which product or service is shown, but also on how it is presented. If the inferred target is to reassure hesitant travelers who want flexible booking options, the system may instantiate that target not only by selecting a hotel with free cancellation, but also by generating a personalized sponsored message that emphasizes low-risk booking, or producing a creative that frames the offer around flexibility and trust. In this setting, generation is no longer limited to producing an identifier for retrieval. It is used to realize or generate a semantic target in the form of a concrete message or creative.

This target-to-instantiation view also makes the multi-stakeholder nature of recommender systems more explicit~\cite{DBLP:journals/umuai/AbdollahpouriAB20}. 
For users, semantic targets can lead to exposures better reflecting their current needs or uncertainties, rather than merely presenting the nearest available item. For providers or advertisers, the same target can reveal how an offering should be framed, which user concern should be addressed, or which product attribute should be made more salient. For platforms, target-level information can help allocate inventory, sponsored opportunities, and creative resources more effectively. Ultimately, the system can decide whether a target is best realized through an item, an offer, a message, or a generated creative. When no existing item satisfies the inferred target, this gap itself becomes a signal of unmet demand. Thus, recommender systems evolve from engines for retrieving items into systems for translating semantic objectives into concrete exposures that coordinate information, inventory, and stakeholder value at scale.


\subsection{Evaluation and Industrial Deployment}

As recommendation shifts towards semantic planning, existing evaluation protocols based on the Cranfield paradigm, which assumes fixed test collections and static relevance judgments, become insufficient~\cite{DBLP:conf/emnlp/WangTZWW23}. The gold standard must move from surface-level hits toward \textbf{planning resonance}: the degree to which a system's trajectory of decisions navigates users toward fulfilled intent~\cite{DBLP:conf/naacl/YoonHEM24}. This requires sophisticated "evaluation agents" for user simulation and long-term goal tracking~\cite{DBLP:conf/www/ZhuHS25}. Crucially, this does not replace long-term proxies like user retention; rather, it provides the missing causal link by explaining \textit{why} an exposure sequence succeeded or failed to align with user and stakeholder intent. Realizing this vision requires resolving significant open challenges:

\begin{itemize}
\item \textbf{ID stability}: How should semantic IDs be designed to remain consistent across dynamic catalog updates without breaking downstream planning models?
\item \textbf{Target grounding}: How can the planning layer be constrained to avoid producing targets that are semantically coherent but unsatisfiable by current inventory or platform infrastructure?
\item \textbf{Objective observability}: How can multi-stakeholder planning quality be supervised and evaluated when provider and platform objectives are only partially observable?
\end{itemize}

Despite these hurdles, the industrial implications are profound. The platform evolves from a passive traffic distributor into a central agent for interpreting and aggregating demand. By feeding semantic signals of unmet needs back to providers, platforms enable a deeper strategic coupling that goes beyond short-term conversion. This also opens new deployment directions: semantic targets can serve as a shared interface between ranking, advertising, and content systems, allowing heterogeneous platform components to coordinate around a common exposure objective rather than optimizing independently. In this sense, semantic planning opens a path toward recommender systems where every exposure is a purposeful step toward aligning user intent with ecosystem capabilities.

\section{Conclusion}
One useful way to read the history of recommender systems is through how information is made usable at scale.
For two decades, 
raw IDs provided the necessary anchors for operational stability and exact memorization.
A central trend today is that semantic information is moving from the periphery of item identity closer to the identity layer itself.
This transition, however, should not be viewed as a clean break. We anticipate an era of architectural hybridity:
Raw IDs will remain essential for auditability, system-of-record requirements, and high-fidelity memorization of head items, while semantic IDs may increasingly serve as a structured vocabulary for unifying information and tasks, and for supporting semantic planning under competing stakeholder objectives.
Retrieval-centric designs will remain dominant where stability, interpretability, and strict output control are primary concerns. The core challenge for researchers and practitioners is therefore not to advocate for a single form of identity, but to orchestrate a co-design of semantic structure, planning objectives, and multi-stakeholder objectives.

As recommender systems evolve from passive matching engines to active decision-making systems, the success of the field will be defined by how well we balance the rigid reliability of the past with the generative, intent-driven flexibility of the future.
The move toward semantic IDs is not merely a technical optimization of retrieval. It is also an important step toward semantic planning: a paradigm in which systems first determine what an exposure should accomplish, and only then decide how best to realize it.
\bibliographystyle{ACM-Reference-Format}
\bibliography{ref}

\end{document}